\documentclass[prb,twocolumn,amssymb]{revtex4}
\usepackage{graphicx}
\usepackage{dcolumn}
\usepackage{bm}
\usepackage{wasysym}

\newcommand{\beq}{\begin{equation}}
\newcommand{\eeq}{\end{equation}}
\newcommand{\beqn}{\begin{eqnarray}}
\newcommand{\eeqn}{\end{eqnarray}}

\begin{document}
\title{Liquids in multi-orbital SU($N$) Magnets with Ultracold Alkaline Earth Atoms}
\author{Cenke Xu}
\affiliation{Department of Physics, Harvard University, Cambridge,
MA 02138}

\date{\today}

\begin{abstract}

In this work we study one family of liquid states of $k-$orbital
SU($N$) spin systems, focusing on the case of $k = 2$ which can be
realized by ultracold alkaline earth atoms trapped in optical
lattices, with $N$ as large as 10. Five different algebraic liquid
states with selectively coupled charge, spin and orbital quantum
fluctuations are considered. The algebraic liquid states can be
stabilized with large enough $N$, and the scaling dimension of
physical order parameters is calculated using a systematic $1/N$
expansion. The phase transitions between these liquid states are
also studied, and all the algebraic liquid states discussed in
this work can be obtained from one ``mother" state with
$\mathrm{SU}(2)\times \mathrm{U}(1)$ gauge symmetry.

\end{abstract}
 \maketitle

\section{introduction}

Spin liquid state as an exotic quantum ground state of strongly
correlated systems has been studied for decades
\cite{anderson1973,anderson1987}. Thanks to the active search for
spin liquids in materials in the last few years
\cite{kappa2003,kappa2005}, people are encouraged to believe in
the existence of spin liquids in nature. The stability of spin
liquid usually relies on large number of matter fields which
suppress the continuous gauge field fluctuations. For instance, in
the famous organic salts $\kappa-\mathrm{(ET)_2Cu_2(CN)_3}$
\cite{kappa2003,kappa2005}, one of the proposed candidate spin
liquid involves a spinon fermi surface, where the finite density
of states of matter field tends to suppress the U(1) gauge field
\cite{motrunich2005,sslee2007}. When the spinon fermi sea shrinks
to a Dirac point, one needs to introduce large enough flavor
number ($N_f$) of Dirac fermions to stabilize the spin liquid.
However, large $N_f$ is difficult to realize in SU(2) spin system,
therefore one is motivated to look for systems with large spin
symmetries. Tremendous theoretical and numerical efforts were made
on SU($N$) and Sp($N$) spin systems with large $N$
\cite{sachdev1990,
sachdev1991,coleman2008,coleman2009,Hermele2005a,ran2006,xu2007,xu2008,assaad2005,brad2007}.

It was proposed that spin-3/2 cold atoms can realize Sp(4)
symmetry without fine-tuning \cite{wu2003}. Recently it has been
discovered that an exact SU($N$) spin symmetry with $N$ as large
as 10 can be realized with alkaline earth cold atoms without
fine-tuning any parameter \cite{alkaline}. Because the electrons
carry zero total angular momentum, all the spin components belong
to nuclear spins and hence the interaction between atoms are
totally independent of the spin components, $i.e.$ the system has
SU($N$) symmetry with $N = 2S+1$ for nuclear spin-$S$. Therefore
the alkaline earth cold atom plus optical lattice is a very
promising system to realize the long-sought spin liquids. Besides
the SU($N$) spins, there is another orbital degree of freedom
associated with the alkaline earth atoms, because both the $^1S_0$
and $^3P_0$ orbital levels (denoted as $g$ and $e$ respectively)
have SU($N$) spin symmetry \cite{alkaline}.

Most generally this system has symmetry $\mathrm{SU}(N)_s\times
\mathrm{U}(1)_c \times \mathrm{U}(1)_o$. $\mathrm{U}(1)_c$
corresponds to the conservation of the total atom number $i.e.$
the charge U(1) symmetry; $\mathrm{U}(1)_o$ corresponds to the
conservation of $n_e - n_g$ $i.e.$ the orbital U(1) symmetry. We
will tentatively assume the system has an extra orbital $Z_2$
symmetry corresponding to switching $e$ and $g$ $i.e.$
$\exp(i\frac{\pi}{2}\sigma^x)$, therefore we take the hopping
amplitude of these two orbitals to be equal, also the two
intraorbital Hubbard interactions are equal. Weak violation of
this $Z_2$ symmetry will be discussed in this paper, and we will
show that it is irrelevant to the main physics discussed in this
paper. Under these assumptions, after straightforward algebraic
calculations the Hamiltonian in Eq. 2 of Ref. \cite{alkaline} can
be rewritten as \beqn H &=& \sum_{\langle i,j \rangle \alpha,m} -
t c_{i\alpha m}^\dagger c_{j\alpha m} + H.c. + \sum_i U(n_i -
\bar{n})^2 \cr\cr &+& \sum_a J (T^a_i)^2 + J_z (T^z_i)^2.
\label{ham2}\eeqn $m = 1 \cdots N$, and $\alpha = e, \ g$. Here
$n_i = \sum_{\alpha m}n_{i\alpha m}$ is the total number of the
atoms on each site, $T^a_i =c^\dagger_{i\alpha
m}\sigma^a_{\alpha\beta}c_{i\beta m}$ is the pseudospin vector of
orbital levels. $U$, $J$ and $J_z$ are simple linear combinations
between $U$, $V_{ex}$ and $V$ in Ref. \cite{alkaline}, which are
from the $s-$wave scatterings between atoms. Since different
orbital channels have different scattering lengths, $J$ and $J_z$
terms are allowed to exist because otherwise the system will have
an unphysical SU(2$N$) symmetry. Eq. \ref{ham2} is the starting
point of our study, and since all the fermionic alkaline atoms
under study carry half integer nuclear spins, $N$ will be taken to
be even hereafter.

In order to obtain more solid and quantitative results, we will
keep both orbitals of the atoms at half-filling $i.e.$ $\bar{n} =
N$, $\sum_i n_{i, e} - n_{i, g} = 0$ and put this model on a
honeycomb lattice with only nearest neighbor hopping. Therefore on
top of the global symmetries discussed before, there is another
particle-hole symmetry with $c_{j\alpha m} \rightarrow \eta_j
c^\dagger_{j\alpha m}$, and $\eta_j = 1$ and $-1$ with $j$
belonging to sublattices A and B respectively. If $t$ is the
dominant energy scale of the Hamiltonian, the half-filled fermions
on honeycomb lattice is a semimetal with two Dirac valleys in the
momentum space located at $\vec{Q} = (\pm \frac{4\pi}{3}, 0)$, and
at low energy the band structure can be described by the Dirac
Lagrangian \beqn L = \sum_{a = 1}^{4N} \bar{\psi}_a
\gamma_\mu\partial_\mu\psi_a, \ \ (\gamma_0, \gamma_1, \gamma_2) =
(\tau^z, \tau^y, - \tau^x). \label{dirac}\eeqn The 2$\times$2
Dirac matrices $\tau^i$ are operating on the two sites in each
unit cell on the honeycomb lattice. The Dirac fermion has two
Dirac points at the corners of the Brillouin zone, therefore there
are in total $N_f = 4N$ flavors of 2-component Dirac fermions,
with an enlarged O(8$N$) flavor symmetry at low energy, which will
be manifest after we rewrite the Dirac fermions in terms of
Majorana fermions. The short-range interactions between the Dirac
fermions are irrelevant at the free Dirac fermion fixed point.

In the following we will mostly be focusing on the Mott Insulator
phase of Eq. \ref{ham2} with $U$ dominant. Motivated by the spin
liquid and weak Mott insulator $\kappa-\mathrm{(ET)_2Cu_2(CN)_3}$
\cite{kappa2005,sslee2005}, we want the system close to the Mott
transition so that at short distance it still behaves like a
semimetal, while at long distance the electron $c_{i\alpha m}$
fractionalizes. With $J = J_z = 0$, the existence of a
fractionalized phase close to the Mott transition on the honeycomb
lattice was shown with a slave rotor calculation in Ref.
\cite{sslee2005}, and the fractionalized spinon has the same
meanfield band structure as the Dirac semimetal. In our system
with nonzero $J$ and $J_z$, various strongly correlated liquid
states with coupled spin, charge and orbital fluctuations can be
realized in different parameter regimes of Eq. \ref{ham2}, and all
the liquid states can be obtained from the $\mathrm{U}(1) \times
\mathrm{SU}(2)$ spin liquid that will be studied first.

\section{liquid states}

\subsection{U(1) $\times$ SU(2) spin liquid, the mother state}

\begin{figure}
\includegraphics[width=2.5in]{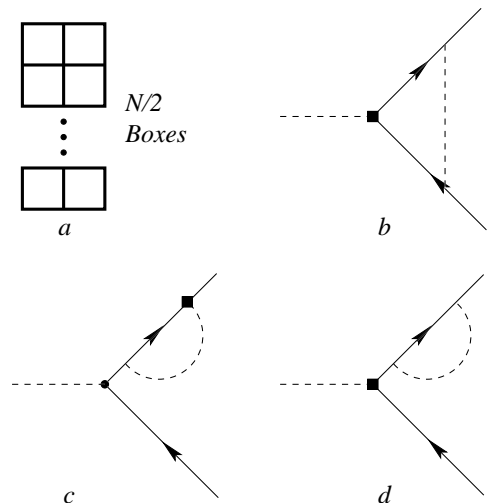}
\caption{$a$, the Young tableau of the representation of $N$
SU($N$) fermions on each site when orbital is constrained to be
$\mathrm{SU}(2)_o$ singlet, $N$ has to be an even number. $b$, $c$
and $d$, Feynman diagrams which contribute to the RG flow of the
velocity anisotropy Eq. \ref{anisotropyRG}, the solid square
stands for the vertex $\sigma^3\gamma_k\partial_k$.}
\label{fdlargen}
\end{figure}

As the first example of liquid state, let us take both $U, \ J$
dominate $t$, while keeping $J_z = 0$ tentatively. In this case
the symmetry of Eq. \ref{ham2} is enhanced to
$\mathrm{SU}(N)_s\times \mathrm{U}(1)_c \times \mathrm{SU}(2)_o$.
When $U$ and $J$ both dominate the kinetic energy, the system
forbids charge fluctuations away from half-filling $n = N$ on each
site, and also forbids orbital-triplet fluctuations, $i.e.$ the
low energy subspace of the Hilbert space only contains orbital
$\mathrm{SU}(2)_o$ singlet. The Young tableau of the SU($N$)
representation on each site has two columns with $N/2$ boxes each
column, which is a large-$N$ generalization of SU(2) spin-1 (Fig.
\ref{fdlargen}$a$). The half-filling constraint on the low energy
Hilbert space implies that one can do a local U(1) rotation on the
fermions, which will be manifested by introducing a U(1) gauge
field $a_\mu$ coupled to the charge degree of freedom as usual.
The orbital-singlet constraint on each site implies that the local
$\mathrm{SU}(2)_o$ transformation will not change the physical
state, and this local invariance can be described by a SU(2) gauge
field coupled to the orbital indices of $c_{i\alpha m}$.

More formally, one can introduce the bosonic U(1) slave rotor
$b_i$ and SU(2) slave rotor, $2\times 2$ matrix field
$h_{\alpha\beta}$, as well as fermionic spinon $f_{i\alpha m}$ as
following \cite{sslee2005}: \beqn c_{i\alpha m} = b_i
h_{\alpha\beta}f_{i \beta m}. \label{decomposition}\eeqn We will
call $b$ the chargeon and $h_{\alpha\beta}$ the triplon field. $h$
is a group element of SU(2), with SU$(2)_L$$\times$SU$(2)_R$
transformation: $h \rightarrow \mathcal{M}_L h\mathcal{M}_R$. The
SU$(2)_L$ symmetry is the physical $\mathrm{SU}(2)_o$ symmetry of
the orbitals, while the SU$(2)_R$ symmetry is a local SU(2) gauge
symmetry, which leaves the physical operator $c_{i\alpha m}$
invariant with an accompanied SU(2) gauge transformation on
$f_{i\alpha m}$: $f \rightarrow \mathcal{M}_R^{-1} f$. The
chargeon $b_i$ grants the spinon $f_{i\alpha m}$ a U(1) gauge
symmetry as usual $b_i \rightarrow b_ie^{i\theta_i}$, $f_{i\alpha
m} \rightarrow f_{i\alpha m}e^{-i\theta_i}$, and $b_i$ also
carries the $\mathrm{U}(1)_c$ charge $i.e.$ $b_i$ will couple to
the external electromagnetic field if the fermions $c_{i\alpha m}$
were electrons. The properties of U(1) and SU(2) slave rotors were
discussed in Ref. \cite{sslee2005} and Ref. \cite{Hermele2007a}
respectively, although the SU(2) slave rotors in Ref.
\cite{Hermele2007a} was engineered from a very different set-up.

The U(1) and SU(2) gauge symmetry can be manifested by
reformulating the hopping term of Eq. \ref{ham2} using the
decomposition of fermion operator Eq. \ref{decomposition}: \beqn H
= \sum_{<i,j>} - t b^\dagger_ib_j f^\dagger_{i\alpha}
h^\dagger_{i\alpha \rho} h_{j\rho\beta}f_{j\beta} + H.c. \eeqn And
spinon $f_{i\alpha m}$ hops effectively in a band structure
described by the following meanfield Hamiltonian \beqn H &=&
\sum_{<i,j>} - t \langle U_{ij,\alpha\beta}\rangle
f^\dagger_{i\alpha}f_{j\beta} + H.c. \cr\cr \langle
U_{ij,\alpha\beta}\rangle &=& \langle b^\dagger_ib_j
h^\dagger_{i\alpha \rho} h_{j\rho\beta}\rangle \eeqn The value of
$\langle U_{ij,\alpha\beta}\rangle $ should be solved
self-consistently. If the self-consistent solution $\langle
U_{ij,\alpha\beta} \rangle \sim \delta_{\alpha\beta}$, the U(1)
and SU(2) symmetries are preserved by this meanfield solution. And
the fluctuation on the meanfield solution is the gauge fields:
$U_{ij,\alpha\beta} \sim \langle U_{ij,\alpha\beta}\rangle
e^{-ia_{ij} - \sum_{l = 1}^3 iA^l_{ij} \tau^l/2}$. The dynamics of
slave rotor $b$ and $h_{\alpha\beta}$ are given by the meanfield
decompositions $ -t b^\dagger_ib_j \langle f^\dagger_{i\alpha}
h^\dagger_{i\alpha \rho} h_{j\rho\beta}f_{j\beta} \rangle$ and $-
t h^\dagger_{i\alpha\rho}h_{j\rho\beta}\langle
b^\dagger_ib_jf^\dagger_{i\alpha}f_{j\beta} \rangle$ respectively.

In the Mott Insulator phase but close to the Mott transition, the
spin model after second order $t/U$ perturbation will be very
complicated. However, there is another self-consistent way of
studying this system. Motivated by the existence of the spinon
fermi sea of weak Mott insulator
$\kappa-\mathrm{(ET)_2Cu_2(CN)_3}$, we assume here the weak Mott
insulator phase is a phase in which the chargeon $b_i$ and triplon
$h$ are both gapped, and the fermionic spinon $f_{i\alpha m}$
fills the same mean field band structure as the original fermions
$c_{i\alpha m}$ in the semimetal phase with $N_f = 4N$ flavors of
2-component Dirac fermions at low energy (Eq. \ref{dirac}), and
then we can check the stability of this state. This spin liquid
state corresponds to a mean field solution $\langle b^\dagger_ib_j
h^\dagger_{i\alpha \rho} h_{j\rho\beta}\rangle =
U\delta_{\alpha\beta}$, which preserves the U(1) and SU(2) gauge
symmetry. After taking into account of the U(1) and SU(2) gauge
fluctuation, the low energy field theory of this spin liquid is
described by the following 2+1d electro-weak theory like
Lagrangian: \beqn \mathcal{L}_{ew} = \sum_{a = 1}^{2N}
\bar{\psi}_a\gamma_{\mu}(\partial_\mu - i a_\mu - \sum_{l =
1}^3iA^l_\mu \frac{\sigma^l}{2})\psi_a + \cdots \label{ew}\eeqn
Here $(\gamma_0, \gamma_1, \gamma_2) = (\tau^z, \tau^y, -
\tau^x)$. $\psi$ is the low energy mode of spinon $f$, and $\psi_1
= e^{i\frac{4\pi}{3} x} f$, $\psi_2 = e^{- i\frac{4\pi}{3} x}
i\tau^y f$. Unlike Eq. \ref{dirac}, in Eq. \ref{ew} each Dirac
fermion $\psi$ is a four component fermion, because it contains
both the Dirac indices and SU(2) gauge indices.

The global symmetry of Eq. \ref{ew} is SU(2$N$), which is a
combined symmetry of SU($N$) spin symmetry and Dirac valley
rotation. $\psi$ transforms nontrivially under translation, space
reflection, rotation, time reversal, and particle-hole
transformation as following: \beqn \mathrm{Tr}_1 &:& x \rightarrow
x + 1, \psi \rightarrow e^{i \frac{4\pi}{3} \mu^z}\psi,\cr\cr
\mathrm{Tr}_2 &:& x \rightarrow x + \frac{1}{2}, y \rightarrow y +
\frac{\sqrt{3}}{2}, \psi \rightarrow
e^{i\frac{2\pi}{3}\mu^z}\psi,\cr\cr \mathrm{T} &:& t \rightarrow
-t, \ \psi \rightarrow \gamma^1\mu^y \psi, \ a_\mu, \rightarrow -
a_\mu, \cr\cr && A^1_\mu, A^3_\mu \rightarrow - A^1_\mu, -A^3_\mu,
\cr\cr \mathrm{P}_{\bar{a}, x} &:& x \rightarrow -x, \ \psi
\rightarrow \gamma^1(\vec{r}_{\bar{a}}\cdot\vec{\mu}) \psi, \ a_1,
A^l_1 \rightarrow -a_1, - A^l_1, \cr\cr \mathrm{P}_y &:& y
\rightarrow -y, \ \psi \rightarrow \gamma^2\mu^z \psi, \ a_2,
A^l_2 \rightarrow -a_2, - A^l_2, \cr\cr \mathrm{PH} &:& c_{j\alpha
m} \rightarrow \eta_j c^\dagger_{j\alpha m}, \ \psi \rightarrow
\gamma^2 \mu^x \psi^\dagger, a_\mu \rightarrow -a_\mu, \ \cr\cr &&
A^1_\mu, A^3_\mu \rightarrow - A^1_\mu, - A^3_\mu, \cr\cr
\mathrm{R}_{2\pi/3} &:& \psi \rightarrow e^{i \frac{2\pi}{3}
\gamma^0}\psi. \label{symmetry}\eeqn $\mu^a$ are three Pauli
matrices that operate on the two Dirac valleys. Notice that the
hexagons of the triangular lattice form a triangular lattice with
three sublattices, and $P_{\bar{a},x}$ is the reflection centered
at sublattice $\bar{a}$ of the three sublattices. Vectors
$\vec{r}_{1} = (0, 1)$, $\vec{r}_{2} = (\frac{\sqrt{3}}{2}, -
\frac{1}{2})$, $\vec{r}_{3} = (- \frac{\sqrt{3}}{2},
-\frac{1}{2})$. In the equation above, transformations of gauge
field components are not shown unless they transform nontrivially.
$\mathrm{R}_{2\pi/3}$ is the hexagon centered rotation by
$2\pi/3$. Notice that time reversal transformation (T) always
comes with a complex conjugate transformation, and hence T only
changes the sign of the SU($N$) as well as $\mathrm{SU}(2)_o$
generators that are antisymmetric and purely imaginary, therefore
the SU($N$) and $\mathrm{SU}(2)_o$ Lie algebras are preserved.

The gauge symmetry and global symmetry together guarantee that
none of the apparently relevant perturbations like fermion
bilinears exists in the Lagrangian Eq. \ref{ew}. When $N$ is large
enough the Lagrangian in Eq. \ref{ew} is a conformal field theory
(CFT). The ellipses in Eq. \ref{ew} include all the gauge
invariant four fermion interaction terms which break the SU(2$N$)
global symmetry down to the symmetries of the microscopic
Hamiltonian Eq. \ref{ham2}. All these four fermion interactions
are irrelevant for large enough $N$.
This CFT fixed point is a pure spin liquid state because both the
charge and orbital fluctuations are forbidden. The scaling
dimension of gauge invariant physical order parameters at this CFT
fixed point can be calculated using a systematic $1/N$ expansion
in a similar way as Ref. \cite{wen2002,Hermele2005a,xu2007}, with
the results: \beqn \Delta_{ew}[\bar{\psi}\psi] = 2 +
\frac{128}{3N\pi^2}, \ \ \Delta_{ew}[\bar{\psi} \mathcal{T}^A_{ew}
\psi] = 2 - \frac{64}{3N\pi^2}. \eeqn Here $\mathcal{T}^A_{ew}$ is
the generator of the SU(2$N$) flavor symmetry. SU(2$N$) current
operators $\bar{\psi}\gamma_\mu \mathcal{T}^A_{ew} \psi$ gain no
anomalous dimension from gauge fluctuations. The order parameters
of many competing orders are classified as fermion bilinears of
this spin liquid states. For instance, the three sublattice
SU($N$) columnar valence bond solid order parameter is
$\bar{\psi}\mu^x\psi$ plus two other degenerate configurations
after translation along $x$ direction. The SU($N$) ferromagnetic
and antiferromagnet order parameter are $\bar{\psi}\gamma_0
T^a\psi$ and $\bar{\psi}\mu^z T^a\psi$, with $a = 1 \cdots N^2 -
1$. We can see that the VBS and the AF order parameters have the
same scaling dimension, and it is smaller than the scaling
dimension of FM order parameter based on the $1/N$ expansion. When
the four fermion interaction is strong enough there is a
transition towards a phase characterized by one of the fermion
bilinear order parameters.

We took $J_z = 0$ at the beginning of this section, but the
algebraic spin liquid discussed here is stable against small
$J_z$, because $J_z$ will not introduce any gauge invariant
relevant perturbation to the field theory Eq. \ref{ew}. For
instance, all the fermion bilinears are ruled out by gauge
symmetries and symmetries in Eq. \ref{symmetry} already. Therefore
a small $J_z$ only renormalizes four fermion terms to Lagrangian
Eq. \ref{ew}. More physically speaking, turning on small $J_z$
will not allow any orbital triplon state in the low energy Hilbert
space, therefore the $\mathrm{U}(1)\times \mathrm{SU}(2)$ gauge
field formalism is still applicable.

For the same reason as the previous paragraph, if we introduce a
small perturbation that breaks the orbital $Z_2$ symmetry
exchanging the two orbital levels, no extra relevant gauge
invariant perturbations on Eq. \ref{ew} can be induced. This is
simply because that the spinon $\psi$ does not carry any physical
orbital charge, therefore a small $Z_2$ symmetry breaking will not
be reflected in the CFT. For instance, in the semimetal phase the
$Z_2$ symmetry breaking will lead to a velocity anisotropy between
two orbitals: $\sum_{<i,j>} \delta t c^\dagger_i \sigma^z c_j$.
Written in terms of fractionalized quantities, this term reads
$\sum_{<i,j>} \delta t b^\dagger_ib_j f^{\dagger}_{i \alpha}
h^\dagger_{i \alpha \mu} \sigma^z_{\mu\nu} h_{i \nu\beta}f_{i
\beta}$, which breaks the global $\mathrm{SU}(2)_o =
\mathrm{SU}(2)_L$ symmetry but still preserves the SU(2) gauge
symmetry. The linear order effect of the $\delta t$ term on the
spinon band structure is proportional to $\langle b^\dagger_ib_j
h^\dagger_{i \alpha \mu} \sigma^z_{\mu\nu} h_{i \nu\beta}\rangle$,
and this expectation value is evaluated in the spin liquid state.
Since the spin liquid state is invariant under the $Z_2$ symmetry
$e^{i\frac{\pi}{2}\sigma^x}$, $\langle b^\dagger_ib_j h^\dagger_{i
\alpha \mu} \sigma^z_{\mu\nu} h_{i \nu\beta}\rangle = 0$, hence at
the linear order the band structure of $f$ is unchanged. In fact,
the velocity anisotropy $\sum_{<i,j>} \delta t c^\dagger_i
\sigma^z c_j$ can lead to the following gauge invariant but
``Lorentz symmetry" breaking coupling in addition to the field
theory Eq. \ref{ew}: \beqn \delta L = \sum_{m = 1}^3 s
\mathrm{Tr}[h^\dagger \sigma^z h \sigma^m ]\bar{\psi}\sigma^m
\gamma_k (\partial_k - ia_k - \sum_{l = 1}^3 iA^l_k
\frac{\sigma^l}{2})\psi, \label{anisotropy2}\eeqn where $k$ is $x$
or $y$. $\mathrm{Tr}[h^\dagger \sigma^z h \sigma^m]$ is odd under
the orbital $Z_2$ transformation $\sigma^z \rightarrow -
\sigma^z$, or $\exp(i\frac{\pi}{2}\sigma^x)$. Therefore as long as
the SU(2) slave rotor $h$ remains gapped, this term only induces
irrelevant term after integrating out the gapped $h$. However, as
we will see in the next section, after the condensation of $h$, an
anisotropic velocity of the spin liquid will be induced, and we
have to evaluate this anisotropy with RG equation.

SU(2) gauge field has been introduced in SU(2) and more generally
Sp(2$N$) spin systems with single orbital
\cite{anderson1988,fradkin1988,wen2002a,Hermele2007a}, but there
the local SU(2) gauge symmetry is a transformation mixing particle
and holes of spinons, and hence there is no extra U(1) gauge field
as in Eq. \ref{ew}. This particle-hole SU(2) gauge symmetry has no
straightforward generalization to larger nonabelian gauge
symmetries. In our case the SU(2) gauge field stems from the
physical orbital degeneracy, and a straightforward generalization
to SU($k$) gauge field with $k-$orbitals can be made, as long as
the Hamiltonian favors a total antisymmetric orbital state. In
this case we can again decompose $c_{i\alpha m}$ as $c_{i\alpha m}
= b_i h_{\alpha\beta}f_{i\beta m}$ with $h\in\mathrm{SU}(k)$. When
$k$ is large the SU($k$) gauge field tends to confine gauge
charges, and controlled calculations are difficult. However, here
SU($k$) gauge field fluctuation corresponds to the constraint of
antisymmetric orbital state, which is analogous to large$-S$ of
SU(2) spin system with antisymmetric orbitals. Therefore the gauge
confined phase could be a semiclassical spin ordered phase.

The credibility of the $\mathrm{U}(1)\times \mathrm{SU}(k)$ gauge
field formalism can be tested in one dimension, where many results
can be obtained exactly. For instance, one of the fixed points of
$k-$orbital SU($N$) spin chain is described by the Wess-Zumino
(WZ) model of SU($N$) group at level $k$ \cite{affleck1986}. At
the $\mathrm{SU}(N)_{k}$ fixed point, the exact scaling dimension
of the Neel order parameter is $\Delta = \frac{N^2-1}{N(N + k)}$.
If we apply the $\mathrm{U}(1)\times \mathrm{SU}(k)$ gauge field
formalism to this spin chain, the first order $1/N$ expansion
gives the scaling dimension of Neel order $\Delta = 1 -
\frac{k}{N}$, which is consistent with the exact result. The
equivalence between WZ theory and the constrained fermion was
proved in Ref. \cite{equivalence}. In one dimensional spin chains,
the WZ fixed point is usually not stable \cite{affleck1988} with
half-filling, in the $\mathrm{U}(1)\times \mathrm{SU}(k)$ gauge
field formalism this instability is due to the relevant Umklapp
four-fermion terms for arbitrarily large $N$. However, in 2+1d all
the four-fermion interactions are irrelevant with large enough
$N$, therefore at the field theory level the spin liquid is more
realistic in 2+1d than 1+1d.

Recently it was proposed that the most general ground state for
SU($N$) Heisenberg magnet with fundamental representation on each
site is a gapped chiral spin liquid \cite{hermele2009}. A chiral
spin liquid can be obtained by spontaneously developing
time-reversal and reflection symmetry breaking fermion gap $
\bar{\psi}\psi $ in the U(1)$\times$SU(2) algebraic spin liquid,
which will lead to the Chern-Simons topological field theory for
the gauge fields.

The $\mathrm{U}(1)\times \mathrm{SU}(2)$ spin liquid state is very
constrained, since both the half-filling constraint and
$\mathrm{SU}(2)_o$ singlet constraint are imposed on each site of
the lattice. In the following we will study several other liquid
states in the same system, which can be obtained from softening
part of the constraints on the $\mathrm{U}(1)\times
\mathrm{SU}(2)$ spin liquid state. Therefore the
$\mathrm{U}(1)\times \mathrm{SU}(2)$ spin liquid state is the
``mother" state of everything else in this paper.

\subsection{U(1) spin-orbital liquid}

Now let us take $U$ large, while keeping $J$ and $J_z$ small. When
$U$ becomes dominant, the system forbids charge fluctuations, but
allows for coupled spin and orbital fluctuations. In this case we
can just introduce chargeon $b_i$ and spinon $f^{(1)}_{i\alpha m}$
as $c_{i \alpha m} = b_i f^{(1)}_{i \alpha m}$ with a local U(1)
gauge symmetry. Here the spinon $f^{(1)}_{i\alpha m}$ is
equivalent to $\sum_{\beta} h_{i\alpha\beta}f_{i\beta m}$ with
$f_{i\beta m}$ defined in the previous section. Therefore the new
spinon does not carry any SU(2) gauge charge, but carries the
physical $\mathrm{SU}(2)_o$ charge. If the fermionic spinon
$f_{i\alpha m}$ fills the same mean field band structure as the
original fermions $c_{i\alpha m}$ $i.e.$ $\langle b^\dagger_i
b_j\rangle$ is a constant, the low energy field theory of this
spin-orbital liquid is described by the following 3D QED
Lagrangian: \beqn \mathcal{L}_{qed} = \sum_{a =
1}^{4N}\bar{\psi}_a\gamma_\mu(\partial_\mu - i a_\mu)\psi_a +
\cdots \label{qed}\eeqn with global flavor symmetry SU(4$N$). The
existence of this phase has been shown with a U(1) slave rotor
meanfield calculation in Ref. \cite{sslee2005}. This type of
Lagrangian has been studied quite extensively in the past, because
several other spin liquid states also have the 3D QED as their low
energy effective field theory \cite{wen2002,xu2007,Hermele2005a}.
It is well-known that when $N_f = 4N$ is larger than a critical
number, the 3D QED is a CFT \cite{confine1990}. Since this CFT
fixed point involves both spin and orbital degrees of freedom (but
no charge fluctuation), we will call this CFT fixed point a U(1)
spin-orbital liquid.

In the well-known staggered flux state of SU(2) spin system, $N_f
= 4$ \cite{wen2002,Hermele2005a,wen2002a}, while in the
spin-orbital liquid states of alkaline earth atoms under study,
$N_f = 4N$ can be as large as 40, therefore it is a much more
promising system to realize this CFT. The first order $1/N_f$
expansion gives us the following results for the scaling
dimensions: \beqn \Delta_{qed}[\bar{\psi}\psi] = 2 +
\frac{32}{3N\pi^2}, \ \ \Delta_{qed}[\bar{\psi}
\mathcal{T}^A_{qed} \psi] = 2 - \frac{16}{3N\pi^2}. \eeqn
$\mathcal{T}^A_{qed}$ is the generator of the SU($N_f$) flavor
symmetry group. Again the SU($N_f$) current $\bar{\psi}\gamma_\mu
\mathcal{T}^A_{qed}\psi$ gains zero anomalous dimension. We can
compare the U(1) gauge field formalism and $1/N$ expansion to the
exact result of SU(2$N$) chains in one dimension, and the $1/N$
expansion gives the exact result as WZ model at level $k = 1$.

Now let us again introduce the $Z_2$ symmetry breaking
perturbation $\sum_{<i,j>} \delta t c^\dagger_i \sigma^z c_j =
\sum_{<i,j>} \delta t b^\dagger_i b_j f^{(1)\dagger}_i \sigma^z
f^{(1)}_j $. Unlike the $\mathrm{U}(1)\times \mathrm{SU}(2)$ spin
liquid discussed in the previous section, since $\langle
b^\dagger_i b_j \rangle \neq 0$, now this $Z_2$ symmetry breaking
will introduce the following gauge invariant perturbation to the
field theory Eq. \ref{qed}, which cannot be absorbed by rescaling
$\psi$: \beqn \delta L = s \bar{\psi}\sigma^3\gamma_k(\partial_k -
ia_k)\psi, \label{anisotropy}\eeqn here $k = x, y$ only includes
the spatial coordinates. Physically this term corresponds to the
velocity difference between the $e$ and $g$ orbitals, and it can
be viewed as Eq. \ref{anisotropy2} after the condensation of SU(2)
slave rotor $h$. We can evaluate the RG flow of this term at the
order of $1/N$ trough Feynman diagrams Fig. \ref{fdlargen}$b$, $c$
and $d$, as was done in Ref. \cite{Hermele2005a}. And the result
is that \beqn \frac{ds}{d\ln l} = - \frac{16}{15\pi^2N}s.
\label{anisotropyRG}\eeqn Therefore this perturbation is
irrelevant under RG flow.

Now if we gradually increase $J$ in Eq. \ref{ham2}, finally the
orbital triplons will be excluded from the low energy Hilbert
space, and the $\mathrm{U}(1)\times \mathrm{SU}(2)$ spin liquid
state discussed in the previous section becomes the candidate
ground state. The phase transition between the U(1)$\times$SU(2)
spin liquid and the U(1) spin-orbital liquid can be driven by
condensing the triplon field $h_{\alpha\beta}$, which can also be
parametrized as $ h = \phi_0 I + i\phi_1 \sigma^1 + i\phi_2
\sigma^2 + i\phi_3 \sigma^3$, $\vec{\phi}$ is a real O(4) vector,
and $\sigma^a$ are Pauli matrices. Further we can define CP(1)
field $z = (z_1, z_2)^t$, and $ z_1 = \phi_0 - i \phi_3$, $z_2 =
\phi_2 - i\phi_1$. Now this phase transition can be described by
the following Lagrangian: \beqn \mathcal{L} = \mathcal{L}_{ew} +
|(\partial_\mu - \sum_{l = 1}^3 iA^l_\mu \frac{\sigma^l}{2})z|^2 +
r_z|z|^2 + \cdots \label{su2higgs}\eeqn with critical point $r =
0$. $\mathcal{L}_{ew}$ is given by Eq. \ref{ew}. After the
condensation of $z$, all three SU(2) gauge field $A^l_\mu$ will be
higgsed and gapped out, and the remnant gauge field is $a_\mu$.
Based on the definition of $f$ in Eq. \ref{decomposition}, the
condensation of $h_{\alpha\beta}$ implies that $f^{(1)}$ and $f$
becomes equivalent after a global SU(2) rotation. This phase
transition is beyond the Landau's theory, because neither side of
the phase transition can be characterized by an order parameter.
For general SU($k$) gauge symmetry with $k
> 2$, condensation of matrix field $h_{\alpha\beta}$ always gaps
out all components of nonabelian gauge fields.

Since the fermion number in $\mathcal{L}_{ew}$ is large, one can
use a systematic $1/N$ expansion to study the universality class
of the transition Eq. \ref{su2higgs}, and the large fermion flavor
number will suppress the SU(2) gauge fluctuations. For instance,
in the large-$N$ limit, we can view the SU(2) gauge field
completely suppressed, then the transition described by Eq.
\ref{su2higgs} belongs to the $3d$ O(4) universality class. Notice
that other gauge invariant and symmetry allowed couplings between
$z$ and $\psi$ are at very high order, and hence are irrelevant at
this transition. For instance, coupling $|z|^2\bar{\psi}\gamma_0
\psi$ violates the particle-hole symmetry. And the $Z_2$ symmetry
breaking term Eq. \ref{anisotropy2} is also irrelevant at this
transition due to its high scaling dimension from power-counting.

The phase transition between the ordinary Dirac semimetal phase
and the U(1) spin-orbital liquid phase can be driven by condensing
the chargeon $b$ in Eq. \ref{decomposition} described by
Lagrangian \beqn \mathcal{L} \sim \mathcal{L}_{qed} +
|(\partial_\mu - i a_\mu)b|^2 + r_b|b|^2 + \cdots
\label{higgs}\eeqn $\mathcal{L}_{qed}$ is given by Eq. \ref{qed}.
The condensation of chargeon $b$ will higgs the U(1) gauge field
$a_\mu$, and release the charge fluctuation from the constrained
Hilbert subspace. In the large-$N$ limit when the gauge field
fluctuation is frozen by fermions, Eq. \ref{higgs} is a $3d$ XY
transition. The velocity anisotropy Eq. \ref{anisotropy} is an
irrelevant perturbation at this transition as well, because the
fluctuation of $b$ will not affect the RG equation Eq.
\ref{anisotropyRG} at the order of $1/N$. A similar metal and weak
Mott insulator transition is studied in Ref.
\cite{senthil2008,senthilkim,balents2009}, where the condensation
of the chargeon rotor $b$ kills the U(1) gauge field fluctuation,
and drive the Mott insulator with spinon into an ordinary metal.

\begin{figure}
\includegraphics[width=3.4in]{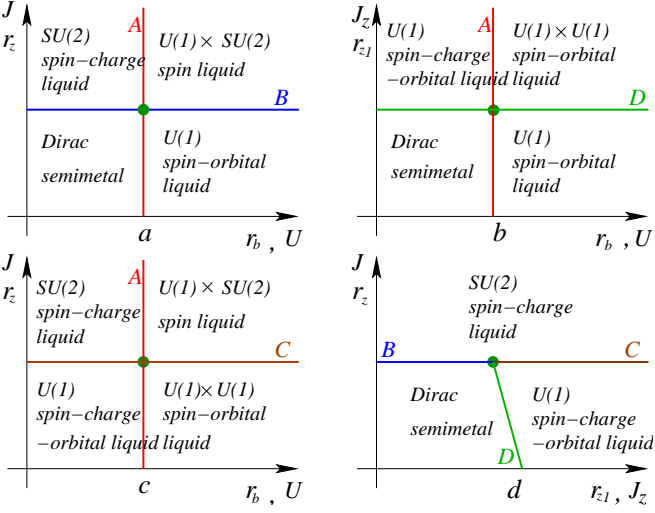}
\caption{Schematic phase diagrams with two tuning parameters. All
the phase transitions denoted as $A$ are described as a Higgs
transition of U(1) slave rotor $b$, such as Eq. \ref{higgs} and
Eq. \ref{ewqcd}; phase transitions denoted as $B$ are Higgs
transition of SU(2) slave rotor $h_{\alpha\beta}$, or CP(1) field
$z$ coupled with SU(2) gauge field, such as Eq. \ref{su2higgs} and
Eq. \ref{qcdmetal}; phase transitions denoted as $C$ are Higgs
transition of SU(2) vector $\vec{\chi}$ coupled with SU(2) gauge
field, for example Eq. \ref{adjoint}. The phase transitions $D$
are Higgs transition of spinon $z_1$ coupled with gauge field
$A^3_\mu$. Notice that in $(a)$, $J_z$ is weak, while in $(c)$
$J_z$ is strong and the constraint $T^z = 0$ is always imposed on
each site. The multi-critical points in these phase diagrams are
discussed in section III. } \label{pdlargen}
\end{figure}

\subsection{SU(2) spin-charge liquid}

The next situation we will consider is to keep $J$ large, and make
$J_z$ and $U$ small. In this case the system forbids triplon
excitations, but charge excitations are allowed. We can introduce
spinon $f^{(2)}$ as $c_{i\alpha m} =
h_{i\alpha\beta}f^{(2)}_{i\beta m}$, and $h_{\alpha\beta}$ is the
same SU(2) slave rotor as introduced in Eq. \ref{decomposition},
while $f^{(2)}_{i\alpha m}$ is equivalent to $b_if_{i\alpha m}$.
This spin-charge liquid state has meanfield solution $\langle
h^\dagger_{i\alpha\rho}h_{j\rho\beta}\rangle \sim
\delta_{\alpha\beta}$, and at low energy can be described by Dirac
fermions coupled with only SU(2) gauge field with a QCD like
Lagrangian \beqn \mathcal{L}_{qcd} = \sum_{a = 1}^{2N}
\bar{\psi}_a\gamma_{\mu}(\partial_\mu - \sum_{l = 1}^3 iA^l_\mu
\frac{\sigma^l}{2})\psi_a + \cdots \label{qcd} \eeqn Since this
state involves both spin and charge excitations, we will call it a
spin-charge liquid. At first glance, the global symmetry in Eq.
\ref{qcd} is $\mathrm{SU}(2N)\times \mathrm{U}(1)$, but the true
symmetry is actually $\mathrm{Sp}(4N)\supset \mathrm{SU}(2N)\times
\mathrm{U}(1)$, and this Sp(4$N$) group is a subgroup of the
O(8$N$) symmetry group of the Dirac fermions in the semimetal
phase without coupling to any gauge field. The enlarged Sp(4$N$)
symmetry was discussed in Ref. \cite{ran2006} in the $\pi-$flux
state of Sp(2$N$) magnets with the same field theory as Eq.
\ref{qcd}.

The Sp(4$N$) symmetry not only contains the explicit SU(2$N$)
flavor symmetry of Eq. \ref{qcd}, but also involves the pairing
channel of $\psi$, because now there is no U(1) gauge field, and
the gauge singlet cooper pairs of $\psi$ are physical operators.
When $k = 2$, The physical order parameters have scaling
dimensions \beqn \Delta_{qcd}[\bar{\psi}\psi] = 2 +
\frac{32}{N\pi^2}, \ \ \Delta_{qcd}[\bar{\psi} \mathcal{T}^A \psi]
= 2 - \frac{16}{N\pi^2}. \eeqn $\mathcal{T}^A \in$ SU(2$N$), and
there are other fermion pairing bilinears with the same scaling
dimension as $\bar{\psi} \mathcal{T}^A \psi$ due to the enlarged
Sp(4$N$) symmetry. The enlarged symmetry is special for $k = 2$,
for QCD Lagrangian with SU($k$) gauge group with $k
> 2$, since it is impossible to form SU($k$) singlet cooper pair,
the global symmetry is the apparent $\mathrm{SU}(2N)\times
\mathrm{U}(1)$ symmetry.

Again the SU(2) spin-charge liquid can be obtained from the
U(1)$\times$SU(2) spin liquid by ``releasing" the charge degree of
freedom, through condensing chargeon field $b_i$ in Eq.
\ref{decomposition}. The Lagrangian is similar to Eq. \ref{higgs}:
\beqn \mathcal{L} \sim \mathcal{L}_{ew} + |(\partial_\mu - i
a_\mu)b|^2 + r_b|b|^2 + \cdots \label{ewqcd}\eeqn After the
condensation of $b$, $f$ and $f^{(2)}$ are identical based on
their definitions. The transition between the SU(2) spin-charge
liquid and the ordinary semimetal phase can be described by
condensing triplon $z_\alpha$ from the SU(2) spin-charge liquid
state, with Lagrangian similar to Eq. \ref{su2higgs}: \beqn
\mathcal{L} \sim \mathcal{L}_{qcd} + |(\partial_\mu - \sum_{l =
1}^3 iA^l_\mu \frac{\sigma^l}{2})z|^2 + r_z|z|^2 + \cdots
\label{qcdmetal}\eeqn After the condensation of the CP(1) field,
physical fermion $c_{i\alpha m}$ and spinon $f^{(2)}_{i\alpha m}$
are identical after a global SU(2) rotation. In the large-$N$
limit, Eq. \ref{ewqcd} and Eq. \ref{qcdmetal} describe a $3d$ XY
and $3d$ O(4) transition respectively. A similar orbital $Z_2$
breaking term is present in the field theory Eq. \ref{qcd} and Eq.
\ref{qcdmetal}, but again this term only leads to irrelevant
effects.

\subsection{U(1) $\times$ U(1) spin-orbital liquid}

If $J$ is small compared with $t$, while both $U$ and $J_z$ are
much larger, then although the charge fluctuation will still be
forbidden, the Hamiltonian gives a green light to one component of
the orbital triplet state: the state $(|e,g\rangle +
|g,e\rangle)/\sqrt{2}$ with $T^z = 0$. Therefore there are two
U(1) constraints on the system: $n_e + n_g = N$, $n_e - n_g = 0$,
therefore we need to introduce spinon which is invariant under
both U(1) charge rotation and orbital rotation generated with
$T^z$. 
Therefore in the proximity of the semimetal phase, the most
natural liquid state with these constraints on the honeycomb
lattice is described by the Lagrangian with two U(1) gauge fields
\beqn \mathcal{L}_{qed2} = \sum_{a =
1}^{2N}\bar{\psi}_a\gamma_\mu(\partial_\mu - i a_\mu - iA^3_\mu
\frac{\sigma^3}{2})\psi_a + \cdots \label{qed2} \eeqn with flavor
symmetry $\mathrm{SU}(2N)_+\times\mathrm{SU}(2N)_- \times Z_2$.
The two $\mathrm{SU}(2N)_{\pm}$ groups are generated by
$\mathcal{T}^A_{\pm} = \mathcal{T}^A_{ew}(1 \pm \sigma^3)/2$
respectively, and the $Z_2$ symmetry exchanges $\pm$. The scaling
dimensions of gauge invariant operators to the first order of
$1/N$ are \beqn \Delta_{qed2}[\bar{\psi}\psi] &=& 2 +
\frac{64}{3N\pi^2}, \ \Delta_{qed2}[\bar{\psi} \mathcal{T}^A_{\pm}
\psi] = 2 - \frac{32}{3N\pi^2}. \eeqn If we turn on the $Z_2$
symmetry breaking perturbation on the lattice, as in the third
section, the following term will be induced in the field theory:
\beqn \delta L = s \bar{\psi}\sigma^3\gamma_k(\partial_k - ia_k
)\psi - \frac{s}{2}iA^3_\mu \bar{\psi}\gamma_k\psi. \eeqn The
scaling dimension of this perturbation can be calculated in the
same way as the third section, and it is still irrelevant
according to the RG equation at the order of $1/N$. In the rest of
this paper this $Z_2$ symmetry breaking will not be mentioned
unless it is relevant.

Since the $\mathrm{U}(1)\times \mathrm{U}(1)$ spin-orbital liquid
only allows one of the orbital triplet states, We can obtain the
$\mathrm{U}(1)\times \mathrm{U}(1)$ spin-orbital liquid by
higgsing two components of the SU(2) gauge field in the
$\mathrm{U}(1)\times \mathrm{SU}(2)$ spin liquid discussed before.
We already showed that condensing a fundamental spinor of the
SU(2) gauge group will gap out all three components of the gauge
fields, but if we just condense an adjoint vector of SU(2) gauge
group, only two of the three components of the gauge field will be
gapped. Therefore starting with the ``mother" state
$\mathrm{U}(1)\times \mathrm{SU}(2)$ spin liquid, the
$\mathrm{U}(1)\times \mathrm{U}(1)$ spin-orbital liquid can be
obtained by condensing SU(2) vector $\vec{\chi} = z^\dagger
\sigma^a z$ instead of $z$ itself. This transition can be
described by the field theory \beqn \mathcal{L} = \mathcal{L}_{ew}
+ \sum_{i = 1}^3 \frac{1}{g}(\partial_\mu \chi_i - \sum_{j,k =
1}^3 \epsilon_{ijk}A^j_\mu\chi_k)^2 + \cdots \label{adjoint} \eeqn
$\epsilon_{ijk}$ is the total antisymmetric tensor, and also the
adjoint representation of SU(2) gauge group: $t^a_{ij} =
i\epsilon_{aij}$. Without loss of generality, we take $\vec{\chi}$
condense along the direction $(0, 0, 1)$, then $A^1_\mu$ and
$A^2_\mu$ are gapped out, while $A^3_\mu$ remains gapless, which
is the same as the $\mathrm{U}(1)\times \mathrm{U}(1)$
spin-orbital liquid. Notice that $\vec{\chi}$ is not a vector of
the physical $\mathrm{SU}(2)_L$ symmetry. To see this explicitly,
we can rewrite $\vec{\chi}$ as \beqn \vec{\chi} =
z^\dagger\vec{\sigma}z \sim \mathrm{Tr}[h^\dagger\sigma^z h
\vec{\sigma}]. \eeqn $h$ is the SU(2) rotor introduced in Eq.
\ref{decomposition}. It is explicit in this equation that
$\vec{\chi}$ is only invariant under the U(1) subgroup generated
by $T^z$, which is the physical
symmetry of the system with finite $J_z$. 
Similarly, if we condense the SU(2) vector $\vec{\chi}_1 \sim
\mathrm{Tr}[h^\dagger\sigma^x h \vec{\sigma}]$ from the mother
state, we would obtain a state with constraint $T^x = 0$ on each
site.

In the large-$N$ limit the SU(2) gauge field is again suppressed
by the fermions, and the transition Eq. \ref{adjoint} is a $3d$
O(3) transition. A similar phase transition was discussed in a
different context \cite{Hermele2007a}. This field theory was also
used as a trial unified theory of electro-weak interaction in
particle physics, and the gapless $A^3_\mu$ was identified as the
photon \cite{glashow1972}. However, nature chooses a different
theory. For larger $k$, condensing adjoint vector of SU($k$) gauge
group always leaves some components of the gauge field gapless.

\subsection{U(1) spin-charge-orbital liquid}

Finally, if we only keep $J_z$ large, while keeping $U$ and $J$
both small, the only constraint on the system is $T^z_j = 0$ on
each site. Then the candidate liquid state in this case is
described by the following lagrangian \beqn L_{qed3} = \sum_{a =
1}^{2N} \bar{\psi}_a\gamma_\mu(\partial_\mu - iA^3_\mu
\frac{\sigma^3}{2})\psi_a + \cdots \label{qed3} \eeqn This state
has spin, charge and two orbital states fluctuation, therefore
following our convention this state will be called U(1)
spin-charge-orbital liquid. This state can be obtained from the
U(1)$\times$U(1) spin-orbital liquid state by condensing chargeon
$b$. Moreover, this U(1) spin-charge liquid state can also be
obtained from condensing SU(2) gauge vector $\vec{\chi}$ in the
SU(2) spin-charge liquid Eq. \ref{qcd}, which gaps out both
$A^1_\mu$ and $A^2_\mu$. In the large-$N$ limit, these two
transitions belong to the $3d$ XY and $3d$ O(3) universality class
respectively.

We can also drive a direct transition between the U(1)
spin-charge-orbital liquid and the semimetal phase, as long as we
can gap out the U(1) gauge field $A^3_\mu$ in Eq. \ref{qed3}. In
the previous paragraph we mentioned that the U(1)
spin-charge-orbital liquid state can be obtained from condensing
vector $\vec{\chi}$ in the SU(2) spin-charge liquid Eq. \ref{qcd}.
After the condensation of $ \vec{ \chi}$, the degeneracy between
the two CP(1) fields $z_1$ and $z_2$ is lifted, due to the gauge
invariance coupling $\vec{\chi}\cdot z^\dagger\vec{\sigma}z$.
Therefore $z_1$ and $z_2$ can condense separately, but not
together. If one of $z_a$ condenses, it will Higgs the gauge field
$A^3_\mu$, and drive the system into the ordinary semimetal phase.
This transition can be described by the field theory \beqn
\mathcal{L} = \mathcal{L}_{qed3} + |(\partial_\mu - i
A^3_\mu)z_1|^2 + r_{z1} |z_1|^2+ \label{qed3metal}\cdots \eeqn In
the large-$N$ limit this transition again belongs to the $3d$ XY
universality class.

\section{Phase diagram and Multi-critical points}

The phase diagram with two tuning parameters $J$ and $U$ and weak
constant $J_z$ is depicted in Fig. \ref{pdlargen}$a$, with four
different liquid phases. And there is a multi-critical point with
both masses of $z_\alpha$ and $b$ vanish. At this multi-critical
point, the field theory reads \beqn \mathcal{L} \sim
\mathcal{L}_{ew} + |(\partial_\mu - i a_\mu)b|^2 + |(\partial_\mu
- \sum_{l = 1}^3 iA^l_\mu \frac{\sigma^l}{2})z|^2 + \cdots
\label{multi1}\eeqn In the large-$N$ limit $z$ and $b$ behave like
a $3d$ O(4) and $3d$ XY transition respectively. On top of this
field theory, the symmetry allows the interaction between $b$ and
$z_\alpha$, such as $|z|^2|b|^2$. It is well-known that at the
$3d$ O(4) and XY transitions, the scaling dimensions
$\Delta[|z|^2]
> \Delta[|b|^2]
> 3/2$ \cite{vicari2004}, therefore this term $|z|^2|b|^2$ is an irrelevant
perturbation in the field theory Eq. \ref{multi1}. Notice that in
principle the velocity of $b$ and $z$ are different, and the
velocities will flow under RG equation with finite $ N$.

Fig. \ref{pdlargen}$c$ is the phase diagram with two tuning
parameters $J$ and $U$ and strong constant $J_z$, where the
orbital constraint $T^z = 0$ is always imposed. Again there is a
multi-critical point with both masses of $z_\alpha$ and $b$
vanish. The field theory at this multi-critical point is \beqn
\mathcal{L} &=& \mathcal{L}_{qcd} + \sum_{i = 1}^3
\frac{1}{g}(\partial_\mu \chi_i - \sum_{j,k = 1}^3
\epsilon_{ijk}A^j_\mu\chi_k)^2 \cr\cr &+& |(\partial_\mu -
ia_\mu)b|^2 + \cdots \eeqn and it is clear that the coupling
$|\vec{\chi}|^2|b|^2$ is irrelevant in the large-$N$ limit due to
the fact that $\Delta[|\vec{\chi}|^2]
> \Delta[|b|^2] > 3/2$.

The multi-critical point in Fig. \ref{pdlargen}$b$ with tuning
parameters $J_z$ and $U$ can be simply described by field theory $
|(\partial_\mu - ia_\mu)b|^2 + |(\partial_\mu - A^3_\mu)z_1|^2$,
and it is stable against interactions between $b$ and $z_1$.
However, the multi-critical point in Fig. \ref{pdlargen}$d$ is no
longer a simple combination between Eq. \ref{su2higgs}, Eq.
\ref{qcdmetal} and Eq. \ref{qed3metal}, because the symmetry of
the system allows the coupling $\vec{\chi}\cdot z^\dagger
\vec{\sigma}z$. The fate of this relevant perturbation is unclear
at this point.

\section{summary and extensions}

In summary, we studied five examples of liquid states motivated by
the orbital flavor and large spin symmetry of alkaline earth cold
atoms. The schematic phase diagrams are depicted in Fig.
\ref{pdlargen}. Experimentally the spin correlation calculated in
this paper can in principle be measured using the momentum density
distribution and noise correlation between atom spins proposed in
Ref. \cite{noise2004}, after releasing the atoms from the trap.
The VBS order which breaks the lattice translation symmetry also
has algebraic correlation in the liquid states. The two atoms
within one valence bond have stronger AF correlation $J \sim
t^2/U$ compared with other atoms, and hence tend to move closer to
each other from the minima of the wells. This super-lattice
structure can also be measured by the density correlation between
atoms, which can be detected by the noise correlation
\cite{demlernote}.

It would also be interesting to test the results of this work by
numerically simulating model Eq. \ref{ham2} as in Ref.
\cite{assaad2005}, since the system is fixed at half-filling, the
sign problem of ordinary interacting fermion system is no longer a
concern. Analytically it is useful to pursue a slave rotor
meanfield calculation like Ref. \cite{sslee2005}. The liquid
states discussed in our paper is expected to occur in a finite
region close to the Mott transition, and by tuning $U$ still
larger there can be a transition from our states into a state with
background nonzero gauge flux through plaquette, or dimerized
valence bond solids \cite{sslee2005}. In our case the interplay
between the U(1) and SU(2) slave rotors make the meanfield
calculation more complicated, and more meanfield variational
parameters need to be introduced. We will study this calculation
in new future.

In the current work, the universality class of all the phase
transitions between different liquid states was only discussed in
the large-$N$ limit. Since the number of boson field at this
transition is not large, the $1/N$ expansion at the transition is
actually nontrivial. Let us take the simplest quantum critical
theory Eq. \ref{higgs} as an example. If there is no gauge field
$a_\mu$, the transition is a $3d$ XY transition, whose critical
exponents can be obtained by summing over the Feynman diagrams to
all orders of an $\epsilon = 4 - d$ expansion. The $1/N$
correction from the gauge field propagator will enter the Feynman
diagram at every order of $\epsilon$ expansion, therefore it is a
nontrivial task trying to sum over all the diagrams at $1/N$
order. However, if we generalize the boson number to large $N_b$,
then a systematic expansion of both $1/N$ and $1/N_b$ can be
straightforwardly carried out, as was studied in Ref.
\cite{sachdevkaul}.

Our formalism can be applied to other multi-orbital magnets,
including transition metal oxides with orbital degeneracy. We will
explore this possibility in future.

\begin{acknowledgments}

The author thanks M. Hermele and S. Sachdev for helpful
discussions. This work is supported by the Society of Fellows and
Milton Funds of Harvard University, and NSF grant DMR-0757145.

\end{acknowledgments}

\bibliography{sptransition1}
\end{document}